\documentclass[useAMS,usenatbib]{mn2e}

\usepackage{epsfig}

\def\swift{{\it Swift}}

\def\H0{{\rm ~km~s^{-1}~Mpc^{-1}}}

\def\HI{\mbox{H\,{\sc i}}}

\newcommand{\nh}{\ensuremath{{N}_\mathrm{H}}}

\def\lsim{\mathrel{\rlap{\lower3pt\hbox{\hskip0.5pt$\sim$}}
    \raise1pt\hbox{$<$}}}                
\def\gsim{\mathrel{\rlap{\lower3pt\hbox{\hskip0.5pt$\sim$}}
    \raise1pt\hbox{$>$}}}                

\title[The extreme, red afterglow of GRB~060923A]
{The extreme, red afterglow of GRB~060923A: Distance or dust?}
\author[N. R. Tanvir et al.]
{N. R. Tanvir$^{1}$\thanks{E-mail: nrt3@star.le.ac.uk}, 
A. J. Levan$^{2}$,
E. Rol$^{1}$,
R. L. C. Starling$^{1}$,
J. Gorosabel$^{3}$,
\newauthor
R. S. Priddey$^{4}$, 
D. Malesani$^{5}$,
P. Jakobsson$^{4}$,
P. T. O'Brien$^{1}$,
A. O. Jaunsen$^6$,
\newauthor
J. Hjorth$^{5}$, 
J. P. U. Fynbo$^{5}$, 
A. Melandri$^{7}$, 
A. Gomboc$^{8}$, 
B. Milvang-Jensen$^{5}$,
\newauthor
A. S. Fruchter$^{9}$,
M. Jarvis$^{4}$, 
C. A. C. Fernandes$^{10}$, 
T. Wold$^{11}$ \\
$^{1}${Department of Physics and Astronomy, University of Leicester, University Road,
Leicester, LE1 7RH, UK}\\
$^{2}${Department of Physics, University of
  Warwick, Coventry, CV4 7AL, UK}\\
$^{3}${Instituto de Astrofisica de Andalucia (IAA-CSIC),
Camino Bajo de Huetor 50, E-18008 Granada, Spain}\\
$^{4}${Centre for Astrophysics Research, University of
  Hertfordshire, Hatfield, AL10 9AB, UK}\\
$^{5}${Dark Cosmology Centre, Niels Bohr Institute,
University of Copenhagen, Juliane Maries vej 30, 2100, Copenhagen, Denmark}\\
$^6$Institute of Theoretical Astrophysics, University of Oslo, PO Box 1029 Blindern,
N-0315 Oslo, Norway\\
$^{7}${Astrophysics Research Institute, Liverpool John Moores University,
Twelve Quays House, Birkenhead, CH41 1LD, UK}\\
$^{8}${Faculty of Mathematics and Physics, University of Ljubljana,
Jadranska 19, 1000 Ljubljana, Slovenia}\\
$^{9}${Space Telescope Science Institute, 3700 San Martin Drive, Baltimore, MD 21218}\\
$^{10}${Department of Physics, University of Oxford, Keble Road, Oxford, OX1 3RH, UK}\\
$^{11}${Joint Astronomy Centre, 660 N, A'ohoku Place, University Park, Hilo, HI 96720, USA}
}
\begin{document}

\date{Accepted . Received ; in original form}

\pagerange{\pageref{firstpage}--\pageref{lastpage}} \pubyear{2007}

\maketitle

\label{firstpage}

\begin{abstract}
Gamma-ray bursts are powerful probes
of the early universe, but locating and identifying
very distant GRBs remains challenging.
We report here the discovery of the 
$K$-band afterglow
of \swift\ GRB~060923A, imaged within the first hour post-burst, 
and the faintest so far found.
It was not detected in any
bluer bands to deep limits, making it a
candidate 
very high redshift burst ($z\gsim11$).  
However, our later-time optical
imaging and spectroscopy reveal 
a faint galaxy coincident with the GRB
position which, if it is the host, implies a more
moderate redshift (most likely 
$z\lsim2.8$) and therefore that dust
is the likely cause of the very red afterglow colour.
This being the case, it is one of the few instances so far found of a 
GRB afterglow with high dust extinction.
\end{abstract}

\begin{keywords}
gamma-rays: bursts; galaxies: high-redshift
\end{keywords}

\section{Introduction}
\label{intro}

The immense luminosity of gamma-ray bursts (GRBs) means that in
principle they could be observed to very high redshift \citep[e.g.]{Lamb}.  
They are associated with massive stars
\citep[e.g.][]{Hjorth}, and there is accumulating evidence,
albeit indirect, that they are preferentially (although not exclusively)
found in lower-metallicity
environments 
\citep[e.g.][]{Fynbo03,Lefloch03,Tanvir04,Fruchter06,Stanek06}.
It therefore seems likely that
GRBs will be produced even by the earliest generations of population
II stars in the universe, and possibly by population III.  This opens
up the possibility that they could pinpoint the first luminous objects
to form after the Big Bang  \citep[e.g.][]{Bromm06}.  Various authors
have attempted to model early star formation and infer a likely GRB
rate as a function of cosmic time.  Although there are many
uncertainties in these models (which get larger as one attempts to
push them earlier), they do predict that as many as $\sim$5--10\% of
\swift\ GRBs could be at $z>5$  \citep[e.g.][]{Natarajan,Yoon,Bromm06}, 
in broad agreement with observational findings to date
\citep{Jakobsson06,Tanvir07}, 
and help make the case that occasional GRBs could be found at much
higher redshifts.

Candidate high redshift bursts have normally been identified
photometrically by a sharp break in the optical to near-infrared (nIR) spectral energy
distribution (SED) of their afterglow light.  The presence of a strong break
in the afterglow spectrum is indicative of high opacity due to
Ly$\alpha$ absorption from neutral hydrogen clouds in the early universe.
Indeed this technique has already facilitated the identification of
the high redshift afterglows of GRB 050814 \citep{Jakobsson06} and GRB
050904 \citep{Haislip}, the latter of which was confirmed to be at
$z=6.3$ by a deep optical spectrum \citep{Kawai}.  Recently, GRB
060927 was found at $z\approx5.5$, and again its very red $R-I$ colour
provided the first indication of extreme distance \citep{RV07}.

However, the situation can at times be more complicated. Although GRB
afterglows generally produce clean  power-law spectra, they are subject to
reddening from interstellar dust, the imprint of which can mimic the 
Ly$\alpha$ break.  This is especially problematic
if the SED is poorly sampled, as is often the case  due to
the difficulty of obtaining deep simultaneous 
multi-band optical and nIR photometry 
for many bursts, even when well placed for observation. 
Hence, there may be some degeneracy
between a burst that lies at high redshift, and one that is highly
reddened by intervening dust.  
A large proportion of star-formation in
the high-redshift universe is thought to be heavily
dust obscured \citep[e.g.][]{Chapman05,Reddy07}, 
leading to the expectation that
many GRBs, having short-lived progenitors, 
should occur in such regions.  
Rather surprisingly, in practice,
examples of even moderately high extinction are rather rare \citep{Klose00,Levan06a}.
This is
a reflection of the fact that GRB hosts on the average seem to be less dusty than
other populations of high-$z$ star-forming galaxies \citep{Lefloch03,Tanvir04,Lefloch06}, 
and probably also
because GRBs destroy dust for considerable distances along our
line-of-sight to them \citep[e.g.][]{Waxman00,Fruchter01,Savaglio03}.
The observational bias against discovering faint, dust reddened bursts
is doubtless also a contributing factor at some level.
Indeed, \citet{Fiore} have argued that the higher average excess hydrogen
column (\nh) (and
hence presumably obscuration) inferred
in bright \swift\ bursts compared to BeppoSAX and HETE2, may be responsible
for a dearth of $z<2$ in the \swift\ sample with measured redshifts \citep[see also][]{Schady}.

In any event, the potential of GRBs as high redshift probes has encouraged
increasing efforts to obtain
early nIR followup of afterglows.  Many GRBs have had no
optical afterglow found, but only in a few cases were searches 
deep and early enough to make them interesting. 
Here we present deep optical and nIR observations of the afterglow
of GRB 060923A. The nIR observations uncovered a very faint
$K$-band afterglow only an hour after the burst, which was not visible in 
subsequent $J$ and $H$ band imaging. Deep Gemini observations show
that the afterglow was also very faint in the optical. We discuss 
possible explanations of this extreme colour in context of both a
high redshift and high extinction origin for the burst.

\section{Observations}
\label{observations}

\subsection{\swift\ observations} 
\label{SwiftObs}

\begin{figure}
\epsfig{figure=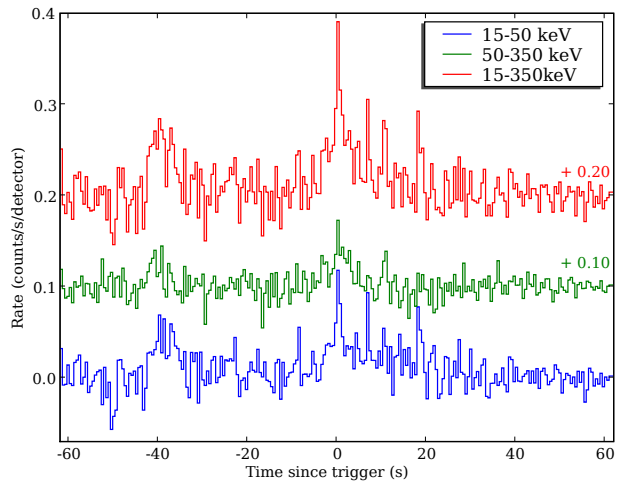,width=9.5cm}
\caption{
{\swift}/BAT light curve of the prompt emission,
sampled at 0.5~s binning.
The lower trace shows softer photon energies (15--50~keV),
the middle trace harder (50--350~keV; displaced vertically
by 0.1 counts~s$^{-1}$~detector$^{-1}$ for clarity), and the top trace
the sum (displaced by 0.2 counts~s$^{-1}$~detector$^{-1}$).
}
\label{BATLC}
\end{figure}

GRB 060923A was discovered by \swift\ at 05:12:15 UT on 23rd September
2006 \citep{Stamatikos}. The burst was of the long-duration class 
with $T_{90} = 51 \pm 1$~s, 
and a photon index $\Gamma = 1.69 \pm 0.23$. The total
fluence of the burst was $(8.7 \pm 1.3) \times 10^{-7}$ ergs cm$^{-2}$
in the 15--150 keV band, and peak photon flux
in a 1~s bin was $1.3\pm0.3$~cm$^{-2}$~s$^{-1}$ \citep{Tueller06}.
The prompt gamma-ray light curve is shown in Fig.~\ref{BATLC}.

Observations with the \swift\ X-Ray Telescope (XRT)
began after 81 seconds and revealed an X-ray afterglow at a location
of $\alpha=$16:58:28.2, $\delta=$+12:21:40.0 (J2000) with
uncertainty 6 arcsec \citep{Conciatore}. The X-ray afterglow
initially declined rapidly with $\alpha = 2.7 \pm 0.3$ 
(following the convention for 
flux density, $F_{\nu}\propto t^{-\alpha}\nu^{-\beta}$; and 90\% 
errors) and then entered a
plateau phase for
approximately 1 hour, finally
settling into a shallower ($\alpha = 1.23 \pm 0.1$) decay. 
Fitting an absorbed power-law spectrum to the orbit 2 (1.2 to 1.9 hours post burst)
and orbit 3 (2.8 to 3.4 hours) data
we infer a spectral slope $\beta = 1.1 \pm 0.2$ (90\% errors) and
an effective column density of neutral hydrogen of
$\nh=14.5^{+3.9}_{-3.5}\times10^{20}$~cm$^{-2}$ ($1\sigma$ errors and Milky Way 
abundance assumed), 
compared to a foreground
estimated as 
${\rm N}_{\rm H,Gal} =
4.52\times 10^{20}$ cm$^{-2}$ \citep{Kalberla}. 
No counterpart was seen with the UV-Optical Telescope (UVOT) to an unfiltered
magnitude limit of about 18.5.

\subsection{Optical and IR observations}
\label{GroundObs}

Our first optical observations were obtained with the 2~m robotic Faulkes
Telescope North (FTN) on Haleakala, starting 
135~s post-burst and
did not detect any counterpart to $R=19.9$, in a 30~s exposure.
Similarly, the SuperLOTIS group reported a limit
of $R=18.4$ in a $5\times10$~s exposure beginning 
at 41~s post-burst \citep{Williams}.
Longer FTN observations over the next hour found deeper limits
of around $R=22.2$.

We obtained our first $K$-band image at the
3.9~m United Kingdom Infrared Telescope
(UKIRT) on Mauna Kea at 06:07 UT, approximately
55 minutes after the burst, which was followed with a 
set of multicolour ($JHK$) observations that began at 06:35 UT. 
These observations were reduced via the
ORAC-DR pipeline \citep[see][]{Cavanagh}. 
Photometric calibration (also for the other nIR observations
reported below) was performed relative to three
2MASS stars in the field.
In these images we found a fading source in the $K$-band
close to the X-ray centroid at position 
$\alpha=$16:58:28.16, $\delta=$+12:21:38.9
(J2000;
astrometric calibration tied to NOMAD 1.0 stars and accurate
to 0.25 arcsec in each coordinate),
which was not visible in either the $J$ or $H$ band observations. 
At $K\approx19.7$ within an hour of the burst, this is 
considerably fainter than previous afterglow detections in
nIR estimated at the same epoch.  For example,
the faint, red afterglow of 
GRB 050215B was also only seen in $K$,
but would have been $K\sim19$ at 1 hour
based on extrapolating back the later time
decay curve \citep{levan06b}.

More optical observations were made at Gemini North on Mauna Kea 
beginning at
07:19 UT (2.1 hours post burst) in the Sloan $r^{\prime}$ and
$i^{\prime}$ filters. 
The images were reduced using standard
IRAF routines,  and calibrated relative to 
a sequence of SDSS stars in the field \citep{SDSS}.
Formally these data yielded no significant
detection to deep limits, although there was 
marginal evidence of excess flux in the $r^{\prime}$ band.
For such a faint afterglow, nIR spectroscopy was not a
feasible option, and further information could only
be gained from imaging.
We therefore obtained additional
observations of the burst position at the Very Large Telescope
(VLT) on Cerro Paranal, in the optical using the FORS1 and FORS2 instruments and 
in the near-IR using ISAAC,
over the following two nights. 
(The location of the burst was such that it was only observable
above 30$^{\circ}$ altitude for about 90 minutes 
after the end of evening twilight
for the northern observatories, and less in the south).
The optical observations
were again reduced via standard procedures using IRAF tools
(in this case the SDSS sequence magnitudes were transformed to Cousins
passbands via the equations of Lupton 
(2005)\footnote{http://www.sdss.org/dr6/algorithms/sdssUBVRITransform.html}, 
to provide a better
match to the filters used),
while ISAAC 
observations were processed with 
{\it eclipse}\footnote{http://www.eso.org/projects/aot/eclipse/}. 
These images provided a clearer detection of a faint, 
extended galaxy coincident with the afterglow position,
the implications of which we discuss below.
Finally we obtained, in better conditions, very late-time ($>$6 months post-burst)
FORS2 $R$-band and ISAAC $K$-band images, and a 40 minute FORS1 optical
spectrum, as
part of the {\swift}/VLT Large Programme on GRB hosts (PI: J. Hjorth).
The former of these confirmed the host detection at $R\approx25.6$, but also
showed the burst to have occurred on the optically brightest
region of the galaxy.  
A very faint continuum trace was seen in the optical spectrum extending
at least as blue as 4600~\AA, but there were no indications of any significant
emission lines.
The two (early- and late-time) ISAAC $K$-band
images each showed marginal ($2\sigma$) evidence of excess
flux in an aperture centred on the afterglow location,
and combining the two measures
yields an estimated host magnitude of $K=21.7\pm0.3$.

\begin{figure}
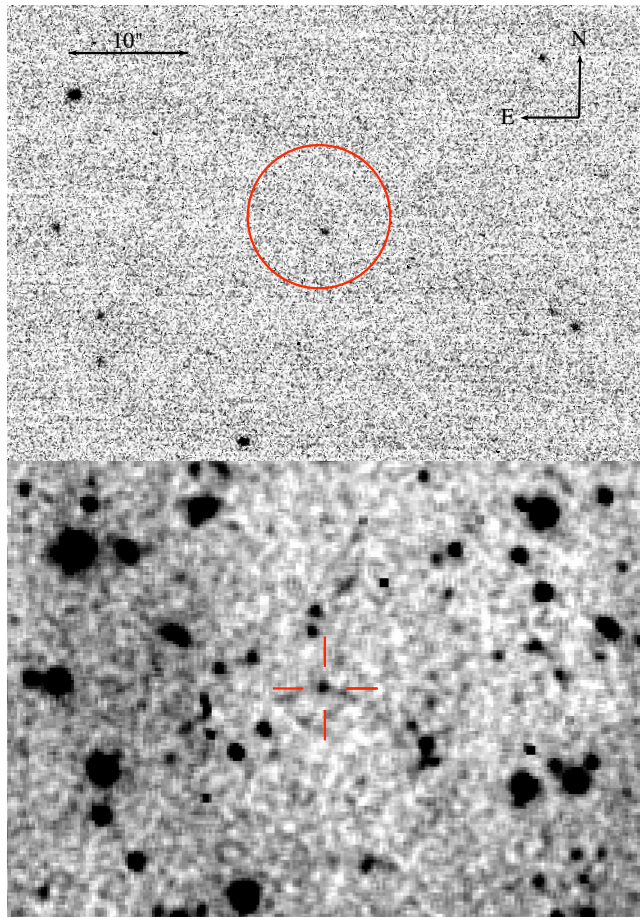

\begin{center}
\epsfig{figure=ukirtk5.epsi,width=8.4cm}
\epsfig{figure=vltr5.epsi,width=8.4cm}
\caption{
Upper: UKIRT/UFTI discovery image of the afterglow of GRB~060923A
in the $K$-band within  
the (plotted) XRT error circle; Lower: final VLT/FORS2 $R$-band image (lightly smoothed), obtained
6 months post-burst,  showing the
faint, extended host galaxy at the same location.  The crosshairs indicate
exact position of afterglow, showing it to be coincident with the brightest
region of the host.  
}
\label{MosaicImage}
\end{center}
\end{figure}

\begin{figure}
\begin{center}
\epsfig{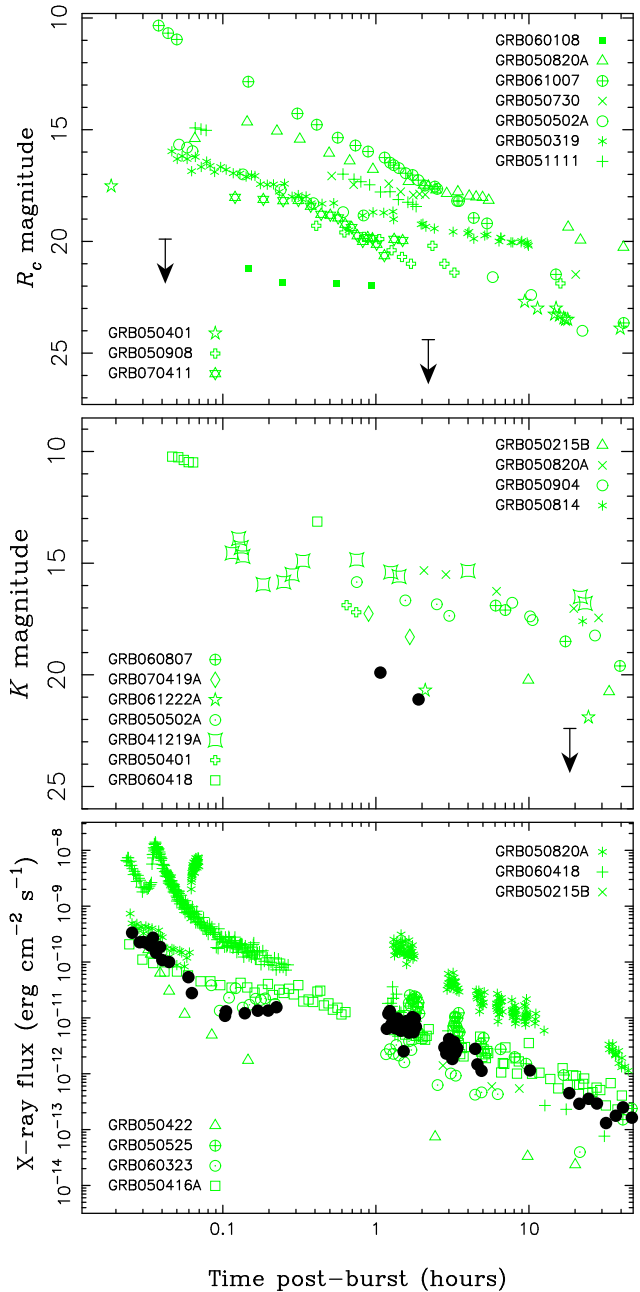}
\caption{
Afterglow light curves in the optical (R-band), near
infrared (K-band) and X-ray (0.3--10~keV) for
GRB 060923A in bold, compared to a set of
other \swift\ GRB afterglows.
These comparison light curves were chosen simply
to illustrate the range in luminosity of typical
\swift\ afterglows, which covers a band of width around 
2 orders of
magnitude in each case.
Unfortunately,
rather few bursts have dense coverage in both
optical and nIR, and only GRB~050820A and GRB~050401 are included
in all three panels.
Note that error bars have been omitted for clarity.
\citep[Data taken from][] {Blake05,Cenko06,CenkoFox,Durig05,Evans,
Guidorzi05,Guidorzi07,Haislip,Huang07,
Jakobsson06,Kann07,
levan06b,Oates06,
Pandey06,RolGCN,Rykoff,
Sharapov05,Quimby06,Watson06,Yost06}.
}
\label{LCfigure}
\end{center}
\end{figure}

The discovery $K$-band image, and final VLT
optical image showing the faint host, are shown
in Fig.~\ref{MosaicImage}.
A log of observations and photometry obtained by
us at the afterglow
position is shown in Table 1. These
data-points and upper limits in optical, infrared
and X-ray are shown in Fig.~\ref{LCfigure},
together with data for a number of other bursts,
illustrating the range of typical afterglow luminosities.

In addition, the GRB location was also observed
by \citet{Fox06} at Gemini North and Keck-I
who found the afterglow to have a brightness of 
$K=20.6$ at $\sim$ 2 hours after the burst. The implied power-law
decay slope to this point
from our first UKIRT observations 
is $\alpha \approx 1.3$, consistent with the decay observed in the X-ray.
They also obtained a deep $J$-band limit of $J>23.7$ at 90 minutes
post burst,
confirming that the afterglow was red even in nIR colours.

These limits certainly make GRB~060923A one of the optically
darkest afterglows observed to date.
Other bursts with famously deep non-detections include GRB~970828,
which was undetected to $R=23.8$ at 3.4 hours
\citep{Groot98}, and GRB~050412, with a limit 
estimated at $R=24.9$ at
2.3 hours \citep{Kosugi05}. 
In terms of the X-ray to optical spectral slope, $\beta_{RX}<0.1$,
it is also one of the most extreme bursts detected to date,
requiring a significant spectral break between X-ray and optical,
and therefore ``dark" by this criterion too \citep{Jakobsson04,Rol05}.

\begin{table*}
\label{obslog}
 \centering
 \begin{minipage}{140mm}
  \caption{Log of our ground-based observations of GRB 060923A, and measured photometry.}
  \begin{tabular}{lllccl}
  \hline
Telescope/instrument & $\Delta t$ (hours)& Filter/grism & Magnitude     & Measured flux ($\mu$Jy) \\   \hline
Faulkes-N/Hawkcam    &  0.037--0.050       &   R  & $R_c>$19.9    &  $4\pm14$ \\
Faulkes-N/Hawkcam    &  0.037--0.20       &   R  & $R_c>$21.0    &  $2\pm5$ \\
Faulkes-N/Hawkcam    &  0.037--0.99       &   R  & $R_c>$22.2    &  $0.5\pm1.6$ \\
UKIRT/UFTI           &  0.92--1.25  &   K  & $K=19.7\pm$ 0.1 & 8.2$\pm$0.8 \\	     
UKIRT/UFTI           &  1.39--1.44  &   J  & $J>$23.6       & $-1.6\pm1.1$ \\     
UKIRT/UFTI           &  1.46--1.50  &   H  & $H>$20.6      & $2.1\pm1.7$ \\     
UKIRT/UFTI           &  1.50--1.55  &   K  & $K>$20.1      &  $1.6\pm2.2$ \\	     
Gemini-N/GMOS        &  2.13        &   r$^\prime$  & $r^\prime>$24.7      & $0.25\pm0.11$ \\	     
Gemini-N/GMOS        &  2.33        &   i$^\prime$  & $i^\prime>$24.7      & $0.15\pm0.16$ \\     
VLT-UT2(Kueyen)/FORS1           & 18.73        &   I  &    $I_c=24.7\pm0.3$       &  $0.26\pm0.08$\\	     
VLT-UT1(Antu)/ISAAC            & 18.47        &   K$_{\rm s}$ &  $K=21.6\pm0.4$ &  $1.5\pm0.6$ \\ 
VLT-UT1(Antu)/FORS2           & 42.54        &   R & $R_c=25.5\pm0.2$& $0.18\pm0.04$\\		
VLT-UT1(Antu)/FORS2           & 4300        &   R & $R_c=25.65\pm0.12$& $0.16\pm0.02$\\	
VLT-UT1(Antu)/ISAAC            &   4420   &   K$_{\rm s}$ &  $K=21.7\pm0.4$ &  $1.3\pm0.6$ \\
VLT-UT2(Kueyen)/FORS1           & 5160        &   300V  &          &  \\ 
\hline

\hline
\end{tabular}

Photometry of the afterglow of GRB 060923A, all calibrated relative
to field stars (SDSS for optical and 2MASS for infrared).
$\Delta t$ is the time from 23 Sep 2006 05:12:15 UT (to the start of the observation,
where a range is not given).
We have not corrected for foreground dust
extinction, which is estimated at only $A_V=0.18$ \citep{Schlegel},
and therefore is
essentially negligible given other uncertainties.
The upper limits on magnitudes are $2\sigma$.  The early
measurements were made in 0.9 arcsec 
diameter apertures for UKIRT images and 1.7 arcsec diameter apertures 
for the FTN images.
Later time observations, where host galaxy flux may be expected to dominate,
are made in 2 arcsec diameter apertures (centred on the afterglow location); 
this includes the brightest parts
of the host, but may miss some lower surface brightness light.
Formal flux measurements and $1\sigma$ uncertainties
are reported also for non-detections.
NB. see text for constraining observations by other authors.
\end{minipage}
\label{default}
\end{table*}

\section{Discussion}
\label{discussion}

\begin{figure}
\begin{center}
\epsfig{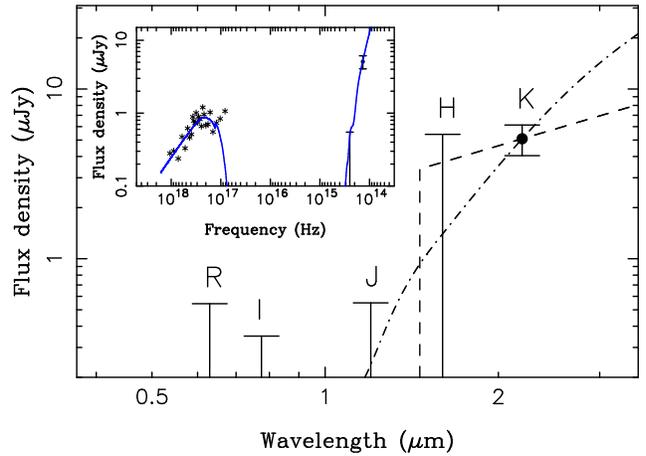}
\caption{
The spectral energy distribution of the afterglow of 
GRB 060923A as observed with
UKIRT and Gemini.
We have subtracted
the host galaxy contribution, as estimated from the later
time observations, and shifted the photometry to a common epoch of 
90~minutes post-burst assuming a decay rate
of $\alpha=1.3$.  Error ranges are $2\sigma$.
Also shown are two illustrative model SEDs: in each
case we assume an intrinsic spectral slope of
$F_\nu\propto\nu^{-1}$, then the dashed line 
uses $z=11$ and no dust extinction,
whereas the dot-dashed line is $z=2$ and a high rest-frame
extinction of $A_V=3$ with a 
\citet{Pei}
SMC extinction law \citep[usually the best fitting to
GRB afterglows, e.g.][]{Kann06,Schady}.
The inset panel shows an absorbed, broken power-law model
($z=2.8$, $A_{V, \rm rest}=2.6$)
which fits both the X-ray and nIR-band data, as described in the text.}
\label{SEDfigure}
\end{center}
\end{figure}

The extremely red colours of the afterglow implied by the
non-detections in the optical suggest that the flux
at wavelengths shortward
of about 1.5~$\mu$m is
significantly suppressed (see Fig.~\ref{SEDfigure}).

Relatively few GRB afterglows exhibit such red colours
\citep{Levan06a}.
This is probably due to a combination of factors:
for instance, GRBs are expected to destroy dust (especially
the smaller grains) for many parsecs along the line of sight; 
and, despite increasingly effective nIR campaigns, there is still some
selection bias against picking up the reddened afterglows.
However, most importantly, studies of GRB host 
galaxies have found relatively few
that exhibit signs of high dust content.  For example,
the number with detectable submm emission is below that
expected based on the proportion of global star formation
thought to take place in ultra-luminous dusty galaxies
\citep{Tanvir04,Michalowski}.  Similarly mm observations at IRAM
\citep[e.g.][]{Priddey}
and mIR observations with 
{\it Spitzer}  \citep{Lefloch06} 
also find that GRB hosts have relatively little
emission on the average.
A small number of GRBs have taken place in extremely
red galaxies (EROs) at moderate redshifts, which is
presumably indicative of dust reddening
(\citealp[e.g. GRB~030115,][]{Levan06a};
\citealp[GRB~020127,][]{Berger07}),
but even these cases are not massive dusty galaxies
judged by their  far-IR emission \citep{Tanvir04,Priddey}.

The rare red afterglows are likely to be the result of
either extreme reddening or high redshift. We discuss each of these
possibilities in relation to GRB~060923A below.

\subsection{The high redshift scenario}

\begin{figure}
\begin{center}
\epsfig{figure=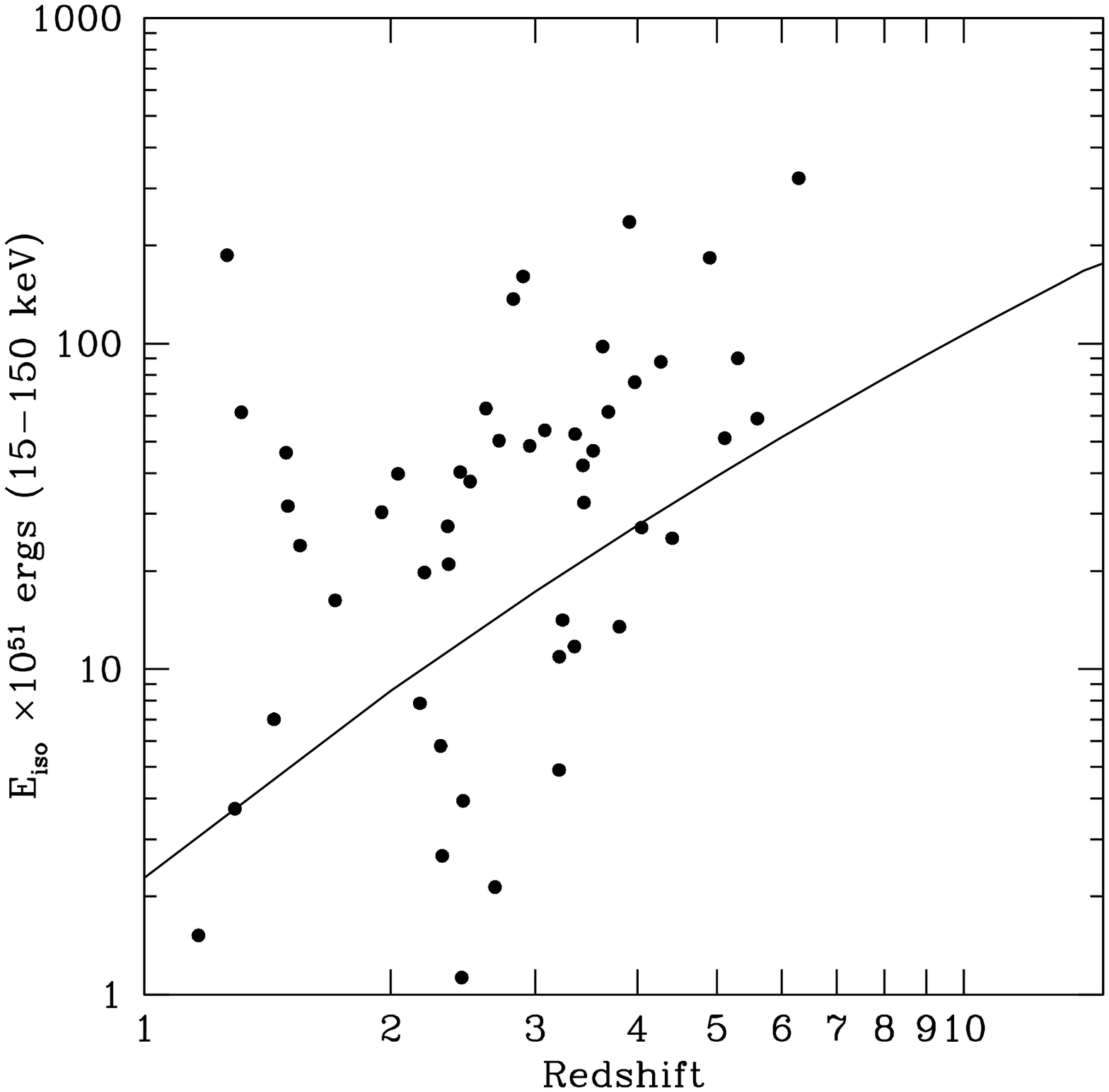,width=8.8cm}
\caption{The intrinsic isotropic
energy of GRB~060923A as would be inferred from the observed 15--150~keV band fluence,
as a function of assumed redshift (solid
line).  The locations of other \swift\ bursts with known redshifts (to September 2007)
are shown as points. As can be seen, although the energy of GRB
060923A would be consistent with the bulk of the GRB population over a
wide range of redshift it would not be overly energetic, even at
$z=10$--15.  We assume a flat, $\Lambda$-dominated cosmology with
$H_0=72$~km~s$^{-1}$Mpc$^{-1}$, $\Omega_M=0.27$, $\Omega_\Lambda=0.73$.}
\label{Eiso}
\end{center}
\end{figure}

As discussed above, highly reddened afterglows and dusty
host galaxies are rare amongst GRBs observed to date.
From Fig.~\ref{Eiso} we see that
the energetics of the burst are such that 
the prompt emission would not be extreme
even if it originated at $z\gsim10$.
The afterglow is faint in all bands, consistent
with a high-$z$ interpretation.
Similarly, the rest-frame duration of the prompt emission
would have been about 5~s at such redshifts,
quite reasonable for a long GRB.

Although the $H$ band limit from our short UKIRT exposure
is not very constraining, it is clear from Fig.~\ref{SEDfigure}
that in the absence of dust extinction, a redshift around $z=11$
or greater is required to explain the red $J-K$ colour
of the afterglow.  As discussed below, however, the
detection of a host galaxy in the optical rules out this
possibility.

\subsection{The high extinction scenario}

The discovery of a host galaxy in the optical 
suggests a moderate redshift, and specifically the detection
in our FORS1 spectrum
of light down to $\lambda\approx4600$~\AA\ 
indicates $z\lsim2.8$, since above that redshift the Ly$\alpha$
forest should render invisible the already weak continuum trace
(a harder upper bound $z<4$ comes from the Lyman limit). 
In fact, the magnitude and colour of the galaxy are
entirely typical of other long-duration GRB hosts \citep{Lefloch03}.
At first sight it may seem somewhat surprising
that the host colour, $R_c-K<4$, is not unusually red 
if the afterglow is highly reddened.   This
could be explained geometrically,
for example if we are viewing the burst through 
an edge-on disk 
or it is embedded in an extended star-forming complex 
where either the dust is patchy and widely distributed, or
its scale length is longer than can be destroyed by the
prompt UV flash from the burst itself.
Interestingly, low surface brightness ``filaments"
can be seen extending south-east and south-west
of the bright knot for several arcseconds (Fig.~\ref{MosaicImage}, lower panel).  This
would be unusually large if it is all part of a single host,
but might indicate a merger or interacting morphology.
This in turn could have 
triggered star formation, some of it dust enshrouded, producing the
bright knot and the GRB progenitor.

However, even prior to identification of the likely host,
there were a number of reasons to doubt
the very high redshift hypothesis.  
In particular, the significant excess column density 
above the Galactic foreground, inferred
from the X-ray spectrum. At very high-$z$ the
rest-frame soft X-rays move out of the {\swift}/XRT
bandpass, so even high columns in the host will
produce little attenuation (presumably exacerbated
by the low metallicity at early cosmic times).
For example, in the sample of 55 \swift\ long-duration
bursts with redshifts considered by \citet{Grupe07},
none of those above $z=2.7$ showed such a high
excess column as seen here.
We should caution, however, that the reliability of column densities
derived from evolving X-ray afterglow spectra
has been questioned by \citet{Butler07}.
This could be particularly relevant if truly at high-$z$
since time-dilation would mean the XRT was observing
early in the rest-frame.
However, the faintness and redness of the afterglow in the
nIR bands, and the fact that the spectral slope
between the nIR and X-ray, $\beta_{KX}\approx0.4$,
also requires a strong spectral break, can all be explained
most easily
if there is significant dust attenuation even in $K$. 
We illustrate this further with the inset panel in Fig.~\ref{SEDfigure},
which shows a combined fit to the X-ray and nIR data.
The model in this case is fixed at $z=2.8$, and assumes
SMC extinction law, metallicity and gas-to-dust ratio.
The unabsorbed spectrum is a $F_{\nu}\propto\nu^{-1}$ power-law
which breaks to $F_{\nu}\propto\nu^{-0.5}$.  In fact this cooling
break must be close to, or in the XRT band to provide a
good simultaneous fit.

\begin{figure}
\begin{center}
\epsfig{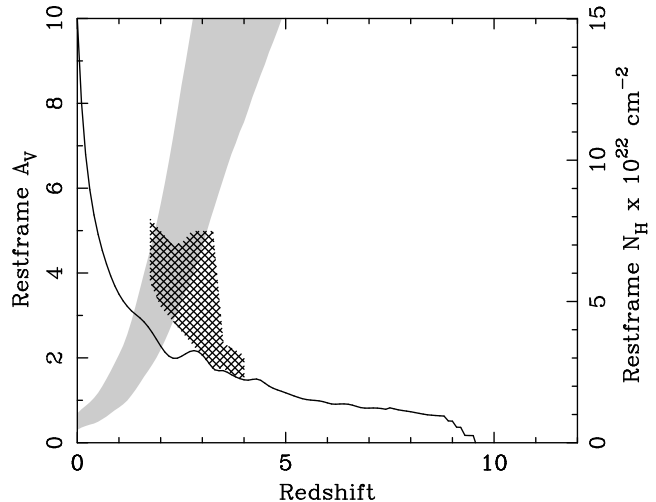}
\caption{
This plot shows various constraints on the combination of
absorption and redshift, derived from the afterglow observations.
As described in the text,
the curve shows the required (minimum) rest frame extinction due to dust 
to explain the observed
$J-K>3.5$ colour of the afterglow of GRB 060923A.  
As with Fig. 4, we  assume a 
\citet{Pei} SMC
extinction law, which should generally be appropriate for \swift\ bursts
at moderate
redshifts,
although perhaps less so at $z>6$ \citep[e.g.][]{Stratta}.
The expected \HI\ column density corresponding
to the $A_V$ scale is shown on the right hand side.
The shaded area then represents the \nh\ column (assuming SMC abundance 
$Z_{\rm SMC}/Z_{\rm MW}=1/8$)
inferred from the curvature of the X-ray spectrum.
Finally, the hatched area shows the results of a combined model fit to the X-ray
and K-band data.
We fixed the Galactic foreground absorption and extinction at $N_{\rm H,Gal} =
4.52\times 10^{20}$ cm$^{-2}$ \citep{Kalberla} and $E(B-V)_{\rm Gal}=0.06$
\citep{Schlegel} respectively. 
}
\label{av}
\end{center}
\end{figure}

More generally, in Fig.~\ref{av} we show how several constraints 
on the afterglow properties would vary with assumed redshift.
For simplicity, and to reduce the number of free parameters,
we assume here an SMC dust extinction law and gas-to-dust ratio
of $N_{\rm H}/E(B-V)\approx4.4\times10^{22}$ atoms~cm$^{-2}$
\citep{Bouchet}.
The solid line shows
the rest-frame
extinction required to produce $J-K=3.5$ from an intrinsic 
$F_{\nu}\propto\nu^{-1}$ afterglow spectrum
\citep[equivalent to $J-K\approx1.7$, e.g.][]{Gorosabel02}.  
In a rather model independent way, then, we can state that
the afterglow should lie around this line or above it.

The shaded region shows the foreground-subtracted \HI\ column
density (the $1\sigma$ full range) derived by
assuming the curvature of the X-ray spectrum is due to absorption by
heavy elements \citep[although we caution that columns inferred this
way usually exceed that measured directly from Ly$\alpha$ absorption;]
[]{Watson07,Jakobsson06b}. For consistency we also assume SMC metallicity
here, so whilst the absolute scale may be wrong, the comparison of $A_V$ and $\nh$ should be valid. 

Finally, the hatched region shows the results of model fitting to the
combined X-ray and K-band data.
Specifically, we extracted a {\swift}/XRT X-ray spectrum such that its observation log
midpoint coincided with the time of the K band photometry. We fitted these
together in count space \citep[e.g.][]{Starling} with models consisting of
an absorbed power law or absorbed broken power law. A single power law is
ruled out from the fits, unless the gas-to-dust ratio is very low.
The broken power law high energy slope was fixed at a spectral index of
$\beta=1.0$ as found in fits to the X-ray spectrum alone. The lower
energy slope was fixed at $\beta=0.5$ as expected for a cooling break 
in the standard fireball model \citep{Sari}.
Below $z\approx1.8$ the fit becomes unacceptably poor ($P<2$\% from a $\chi^2$ test).
The region shown is curtailed at the upper end at $z=4$ based on the limits
from the optical spectrum discussed below.
The vertical extent of the region represents the 90\% error range on $A_V$.
We note that the hatched area is essentially consistent with the requirement
of a reddened afterglow (above the solid curve).
The part of the hatched
region which does not coincide with the grey shaded band corresponds to best
fitting models where the curvature of the X-ray spectrum is partly due to a cooling
break, hence reducing the required absorption.  In fact, the hardness ratio for the orbit 2 spectrum
is consistent with that for orbit 3 and also for the combined spectrum beyond orbit 3,
providing no evidence for a moving spectral break, but the statistics are too 
poor to make a firm statement.

We could, of course, have allowed more freedom in the model fitting,
for example by not fixing the gas-to-dust ratio.
However, allowing this to be a free parameter finds models
with lower \nh\ column at a given extinction,
whereas, for long GRB afterglows, the columns determined from X-rays
are more often in excess of those determined from the reddening \citep[e.g.][]{Starling}.
A larger than expected cooling break ($\Delta\beta>0.5$)
could in principle reduce the required extinction,
but then we would not explain the reddening of the afterglow.
In any case, even with greater freedom it is hard to find any reasonable 
solution at $z<1.5$.

We conclude that the burst likely occurred at $z\gsim1.8$,
which, combined with the constraint $z\lsim2.8$, requires $2\lsim A_V\lsim5$.
This extinction would be high by typical GRB standards, but quite moderate
compared to sight lines close the plane of a disk galaxy or
through large dust-enshrouded star-forming regions.

We note that, whilst not common, 
optically dark bursts in blue galaxies at
moderate redshifts, have been identified before,
for example,
GRB~970828 \citep{Djorgovski01}, GRB~000210 \citep{Gorosabel03},
GRB~051022 \citep{Rol07} and GRB~070306 \citep{Jaunsen08}.
Nevertheless, it remains odd that the location of the burst
in this case
places it within $0.2$~arcsec, corresponding to at most
a kiloparsec or two (at any plausible redshift),  
of the optically brightest part of the host.  This strong spatial coincidence between burst
position and the optically brightest regions of their hosts is seen in many other 
low-reddening long GRBs \citep{Fruchter06}, and may be a
consequence of the very short life-times of massive-star GRB progenitors,
which do not move far from the star-forming region of their birth \citep{Larsson}.
In the case of GRB~060923A, unless there is some separation along the line-of-sight, 
we require the dust-attenuated GRB to be close to
a relatively unreddened region that dominates the optical light of the
galaxy.  A plausible geometry is one in which high optical-depth molecular
clouds provide patchy obscuration of a large star-forming complex 
\citep[cf. NGC 604 in M33 as a possible low-redshift analogue;][]{MA}, and the sight-line to the GRB happens
to intersect one of these.

\section{Summary}
\label{summary}

There is great interest in GRBs with very red optical-nIR colours
since they could be at very high redshift.
GRB 060923A is arguably the most extreme example found to-date,
being detected in the K-band, but with only deep limits in
early observations in all bluer filters.
If purely due to a Ly$\alpha$ break in the near-IR,
this would indicate a redshift beyond $z\sim11$.
However, our  later-time optical observations
revealed a faint galaxy, presumably the host, at the same position, 
which must be at a
more moderate redshift, probably $z\lsim2.8$ but certainly not above $z\approx4$.
The morphology of the host: a bright knot coincident with the
GRB itself, and extended low surface brightness features
may indicate a merger/interaction has produced this burst of star
formation.
A combined analysis of the X-ray and optical/nIR data suggest
that the burst is also likely to have $z\gsim1.8$ in order to reconcile
the absorption required in both bands.

There is, of course, a slim chance that the GRB is coincident with an
unrelated foreground
galaxy.  Following the analysis of \citet{Piro} we calculate a probability that the
afterglow would be found coincident with an unrelated galaxy as bright as $R=25.6$
to be about 3\% -- a small but non-negligible figure.
However given the close alignment of the afterglow with the
brightest knot of the galaxy, and that the colour and magnitude
of the putative
host are otherwise very plausible for a typical 
long-duration GRB, the conservative explanation remains that this is a 
case of dust rather than distance.

GRB~060923A nonetheless is likely to be representative of some proportion
of the dark GRB population which optical surveys are biased against
finding.
It is therefore also  a good example
of the kind of interlopers which we must be able to reject in order
to identify very high redshift GRBs.
This study emphasises that early, deep photometry in a range
of optical and nIR filters is essential to reliably 
identify candidates, and 
followup spectroscopy is highly desirable where
possible.

\section*{Acknowledgments}

We thank the UK Science and Technology Facilities Council for
financial support, in particular NRT for a Senior Research
Fellowship and AJL for a Postdoctoral Fellowship.
The  
research activities of JG are supported by the Spanish Ministry of  
Science through the programmes ESP2005-07714-C03-03 and AYA2004-01515.
DM acknowledges the Instrument Centre for Danish Astrophysics.
PJ acknowledges support by a Marie Curie Intra-European Fellowship 
within the 6th European Community Framework Program under contract 
number MEIF-CT-2006-042001, and a
Grant of Excellence from the Icelandic Research Fund. 
AOJ acknowledges support from the 
Norwegian Research Council, grant 
nr. 166072. 
We also gratefully acknowledge the work of the wider \swift\
team that makes this research possible.

The United Kingdom Infrared Telescope is operated by the Joint
Astronomy Centre on behalf of the U.K. Science and Technology
Facilities Council.

Partly based on observations carried out with the ESO
telescopes under programmes 077.A-0667 and 177.A-0591.

Based on observations obtained at the Gemini Observatory, which is
operated by the Association of Universities for Research in Astronomy,
Inc., under a cooperative agreement with the NSF on behalf of the
Gemini partnership: the National Science Foundation (United States),
the Particle Physics and Astronomy Research Council (United Kingdom),
the National Research Council (Canada), CONICYT (Chile), the
Australian Research Council (Australia), CNPq (Brazil) and CONICET
(Argentina).

The Faulkes Telescopes are operated by the Las Cumbres
Observatory.

The Dark Cosmology Centre is funded by the Danish 
National Research Foundation.

\bsp

\label{lastpage}

\end{document}